\documentclass{article}
\usepackage{spconf,amsmath,graphicx}
\usepackage{braket}
\usepackage{amssymb} 
\usepackage{subcaption} 
\usepackage{setspace}
\usepackage{bbm}
\usepackage{makecell}
\usepackage{booktabs}
\usepackage{arydshln}
\usepackage{xspace}

\usepackage{tabularx}
\usepackage{color}
\usepackage{xcolor}

\title{How Can Quantum Deep Learning Improve Large Language Models?}
\name{Emily Jimin Roh$^\dag$\thanks{This research was supported by the Institute of Information \& Communications Technology Planning \& Evaluation (IITP) grant funded by the Korea government [MSIT (Ministry of Science and ICT (Information and Communications Technology))] (RS-2024-00439803, SW Star Lab) for Quantum AI Empowered Second-Life Platform Technology; and also by the National Research Foundation of Korea (RS-2025-00561377).}, Hyojun Ahn$^\dag$, Samuel Yen-Chi Chen$^\ddag$\thanks{The views expressed in this article are those of the authors and do not represent the views of Wells Fargo. This article is for informational purposes only. Nothing contained in this article should be construed as investment advice. Wells Fargo makes no express or implied warranties and expressly disclaims all legal, tax, and accounting implications related to this article.}, Soohyun Park$^\S$, Joongheon Kim$^\dag$\thanks{Corresponding Author: Joongheon Kim (joongheon@korea.ac.kr)}
}
\address{$^\dag$Korea University \quad $^\ddag$Wells Fargo
\quad $^\S$Sookmyung Women's University}

\begin{document}
\maketitle
\begin{abstract}
The rapid progress of large language models (LLMs) has transformed natural language processing, yet the challenge of efficient adaptation remains unresolved. Full fine-tuning achieves strong performance but imposes prohibitive computational and memory costs. Parameter-efficient fine-tuning (PEFT) strategies, such as low-rank adaptation (LoRA), Prefix tuning, and sparse low-rank adaptation (SoRA), address this issue by reducing trainable parameters while maintaining competitive accuracy. However, these methods often encounter limitations in scalability, stability, and generalization across diverse tasks. Recent advances in quantum deep learning introduce novel opportunities through quantum-inspired encoding and parameterized quantum circuits (PQCs). In particular, the quantum-amplitude embedded adaptation (QAA) framework demonstrates expressive model updates with minimal overhead. This paper presents a systematic survey and comparative analysis of conventional PEFT methods and QAA. The analysis demonstrates trade-offs in convergence, efficiency, and representational capacity, while providing insight into the potential of quantum approaches for future LLM adaptation.
\end{abstract}
\section{Introduction} \label{sec:1}
Large language models (LLMs) have emerged as essential backbones in natural language processing, which makes diverse applications from open-domain dialogue to specialized text generation~\cite{icassp_Die}. Their effectiveness is largely attributed to massive parameterization and extensive pre-training, which provide strong generalization across tasks. Full fine-tuning provides high accuracy but requires extensive resources, limiting deployment in environments with constrained computation or energy budgets~\cite{globecom25HJ}.

To address this limitation, parameter-efficient fine-tuning (PEFT) techniques have emerged as viable alternatives. 
Low-rank adaptation (LoRA) reduces the number of trainable parameters through low-rank decomposition of weight updates~\cite{hu2022lora}. 
Prefix tuning introduces task-specific vectors at the input level, while sparse LoRA (SoRA) extends low-rank approaches with sparsity constraints for improved scalability~\cite{SoRA}. 
Nevertheless, many approaches still require the update of millions of parameters in large-scale models, which imposes significant memory overhead. 

\begin{figure}[t]
    \centering
    \includegraphics[width=0.99\columnwidth]{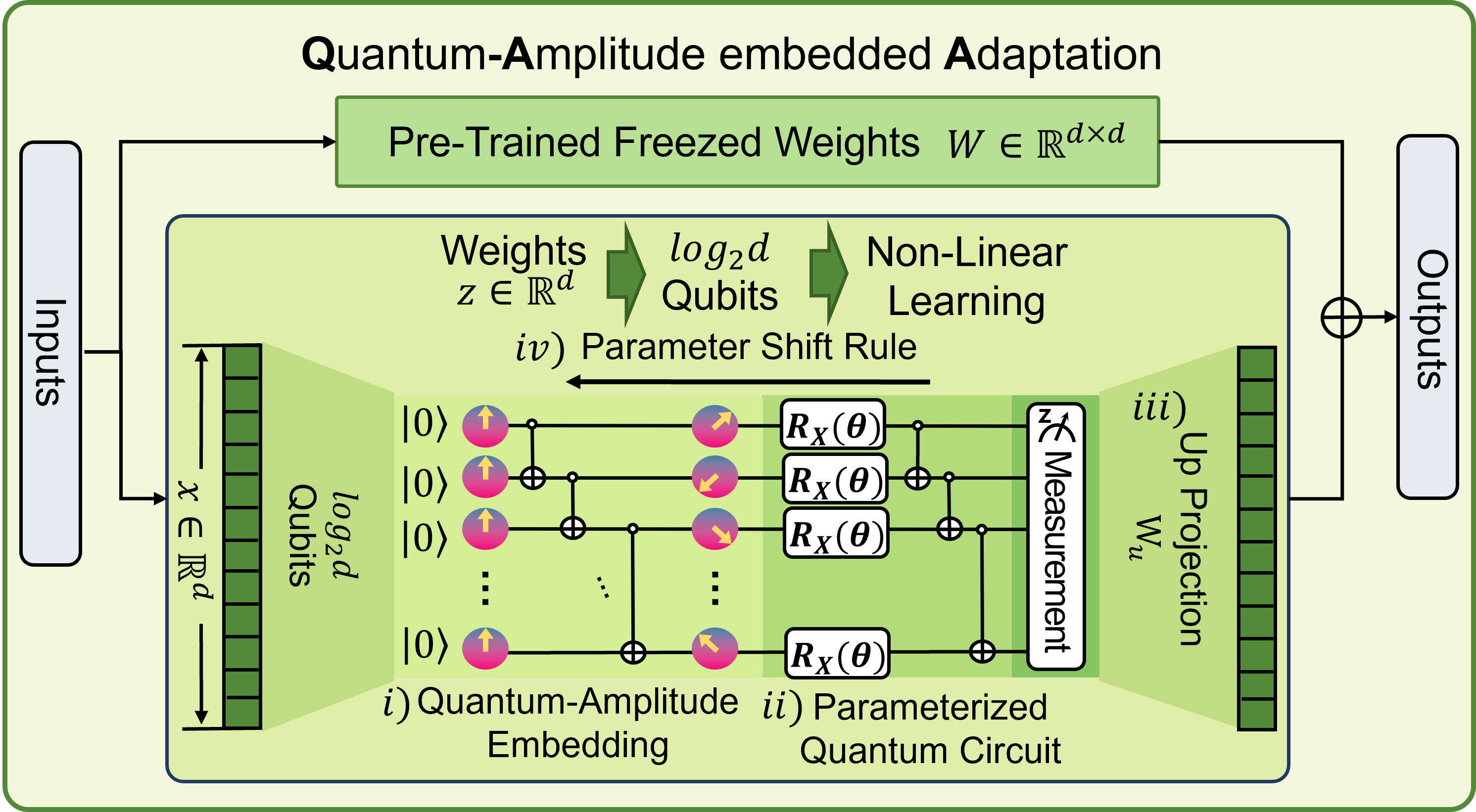}
    \caption{Overview of the QAA framework, a quantum deep learning approach for the LLM fine-tuning.}
    \label{fig:1}
    \vspace{-3mm}
\end{figure}

Quantum deep learning introduces a new paradigm for LLM fine-tuning. Quantum encoding combined with parameterized quantum circuits (PQCs) enables expressive transformations, thereby allowing LLM fine-tuning to be performed more efficiently with a reduced number of trainable parameters~\cite{CM_1}.
Quantum-amplitude embedded adaptation (QAA) extends this principle by mapping classical hidden states into quantum states, which produces compact yet powerful updates~\cite{QAA}. 
Unlike conventional PEFT methods, QAA leverages quantum superposition and entanglement to preserve representational richness under strict parameter constraints. 

\begin{table*}[t]
\centering
\caption{Comparison of representative PEFT methods based on GPT-Neo in terms of main contributions, fine-tuning complexity, and trainable parameter ratio. 
Here, $d$ denotes the hidden dimension size, $r$ is the rank used in low-rank adaptation, $r_{\text{eff}}$ is the effective rank after sparsity adjustment in SoRA, and $l$ the prefix length.}
\renewcommand{\arraystretch}{1.2}
\resizebox{\textwidth}{!}{
\begin{tabular}{c | c | m{8cm} | c | c}
\Xhline{3\arrayrulewidth}
\textbf{Method} & \textbf{Ref.} & \centering\arraybackslash \textbf{Main Contribution} & \textbf{Fine-Tuning Complexity} & \textbf{\#Trainable Parameter and Ratio} \\
\hline
Full Tuning & \cite{Full-Finetuning} & 
Updates all parameters of the pre-trained model without any restriction, which achieves strong downstream performance but with extremely high computational and memory costs. & 
$O(d^2)$ & 125,198,592 (100\%) \\
\hline
LoRA & \cite{LoRA} & 
Introduces trainable low-rank matrices into each layer while freezing the backbone, and this enables efficient adaptation under the hypothesis that model updates are intrinsically low-dimensional. Provides a strong trade-off between performance and efficiency. & 
$O(dr)$ & 147,456 (0.12\%)\\
\hline
SoRA & \cite{SoRA} & 
Extends LoRA by allowing dynamic and sparse adjustment of the intrinsic rank during training. A gating unit, optimized via proximal gradient methods, adaptively prunes redundant components. This achieves higher efficiency and often better accuracy than fixed-rank LoRA while reducing the number of trainable parameters (e.g., 0.91M vs. 1.33M for $r=8$). & 
$O(dr_{\text{eff}}),\ r_{\text{eff}} < r_{\max}$ & 125,337 (0.10\%)\\
\hline
Prefix Tuning & \cite{Prefix_Tuning} & 
Learns a sequence of continuous trainable prefix vectors prepended to the input of each transformer layer. This conditions the model to new tasks without modifying the original weights, but introduces additional sequence length during training and inference. & 
$O(ld)$ &  552,960 (0.44\%)\\
\hline
\textbf{QAA} & \cite{QAA} & 
Proposed quantum-inspired adapter method that leverages amplitude embedding and PQCs. It enables expressive representation power with logarithmic scaling in qubit space, thereby providing parameter-efficient adaptation while maintaining competitive accuracy. & 
$O(d\log d)$ &  \textbf{123,000 (0.09\%)}\\
\Xhline{3\arrayrulewidth}
\end{tabular}
}
\label{tab_peft}
\vspace{-3mm}
\end{table*}

This paper analyzes full tuning, LoRA, Prefix tuning, SoRA, and QAA in the context of LLM adaptation. The discussion provides a evaluation of efficiency and convergence. This study also highlights the unique role of quantum methods in overcoming scalability bottlenecks and shaping the next generation of fine-tuning strategies for LLMs.

\section{Preliminary} \label{sec:2}
\subsection{LLM Fine-Tuning Framework}
Modern LLMs contain billions of parameters, which makes full adaptation prohibitively expensive. 
Given a dataset $\mathcal{D}=\{(x_i,y_i)\}_{i=1}^N$ and a pre-trained model $P_{\Phi}(y \mid x)$ with parameters $\Phi$, 
the full fine-tuning objective can be formulated as,
\begin{equation}
    \max_{\Phi} \sum_{(x,y)\in\mathcal{D}}\nolimits \sum_{t=1}^{|y|}\nolimits \log P_{\Phi}(y_t \mid x, y_{<t}).
    \label{eq:ft}
\end{equation}

Since updating all $\Phi$ is impractical for large $|\Phi|$, PEFT introduces a small set of trainable parameters $\theta$, with $|\theta| \ll |\Phi|$. 
The update function $\Delta h(\theta)$ modifies the model as $\Phi + \Delta h(\theta)$, which is defined as,
\begin{equation}
    \max_{\theta} \sum_{(x,y)\in\mathcal{D}}\nolimits \sum_{t=1}^{|y|}\nolimits \log P_{\Phi + \Delta h(\theta)}(y_t \mid x, y_{<t}).
    \label{eq:peft}
\end{equation}

Beyond classical PEFT, quantum-inspired approaches define $\Delta \Phi(\theta)$ through amplitude embedding and parameterized circuits, enabling compact yet expressive adaptation.

\subsection{Related Work}

Recent studies have proposed a variety of PEFT techniques for adapting LLMs. 
Table~\ref{tab_peft} compares representative approaches in terms of methodological design, fine-tuning complexity, and parameter efficiency. 
Full fine-tuning directly updates the entire parameter set $\Phi \in \mathbb{R}^{|\Phi|}$ of a pre-trained model for each downstream task 
and can be expressed as,
\begin{equation}
    \Phi' = \Phi + \Delta \Phi, \quad \Delta \Phi \in \mathbb{R}^{|\Phi|},
\end{equation}
which achieves strong performance but incurs $O(d^2)$ complexity and requires storing a full model copy per task. 
Here, $d$ denotes the hidden dimension of the model. 
This makes full fine-tuning infeasible for billion-scale LLMs~\cite{Full-Finetuning}.  

To address this limitation, researchers have developed methods that introduce restricted sets of trainable components while keeping the backbone largely frozen. 
LoRA reduces the trainable parameter space by factorizing weight updates into low-rank matrices defined as,
\begin{equation}
    \Delta W = AB^\top, \quad A \in \mathbb{R}^{d \times r}, \; B \in \mathbb{R}^{d \times r}, \; r \ll d,
\end{equation}
and applying the effective weight update $W' = W + \Delta W$, where $W$ denotes the original weight matrix. 
This approach lowers the number of trainable parameters to $O(dr)$ while retaining competitive accuracy~\cite{LoRA}.  

Building on this idea, SoRA extends LoRA by dynamically adjusting and sparsifying the effective rank through a gating vector, 
optimized using proximal gradients as,
\begin{equation}
    \Delta W = A \, \text{diag}(g) B^\top, \quad g \in \mathbb{R}^r,
\end{equation}
which adaptively prunes redundant components. This method often achieves better accuracy than fixed-rank LoRA while using fewer effective parameters~\cite{SoRA}.  

Another approach is Prefix Tuning, which learns continuous prefix vectors $P \in \mathbb{R}^{l \times d}$ that are prepended to the input of each transformer block defined as,
\begin{equation}
    h' = f([P;x]; \Phi),
\end{equation}
where $x$ is the input sequence, $f(\cdot)$ denotes the frozen backbone, and $l$ represents the prefix length. 
The computational cost scales as $O(ld)$~\cite{Prefix_Tuning}.  

More recently, QAA adopts a quantum amplitude embedding strategy that compresses an input $x \in \mathbb{R}^d$ into $\log d$ qubits. 
The embedded states are processed through PQC composed of $R_X$ rotation gates and CNOT entanglement gates, which enable expressive non-linear transformations. 
The output is then mapped back to the original dimension through an additional linear up projection, allowing fine-tuning with a complexity of $O(d \log d)$. 
A more detailed description of QAA is provided in Section~\ref{sec:3}.

\section{Details of Quantum-Amplitude Embedded Adaptation} \label{sec:3}
QAA is presented as a quantum deep learning approach for enhancing the performance of LLMs, where conventional linear adapters are replaced with compact quantum modules that enable expressive and parameter-efficient adaptation.
By embedding hidden states into a quantum Hilbert space, QAA enables non-linear transformations with a logarithmic number of qubits, which produces task-specific residuals $\Delta h$ while significantly reducing parameter counts.

As illustrated in Fig.~\ref{fig:1}, the QAA framework follows four stages: i) quantum amplitude embedding of input activations, ii) quantum processing via PQC, iii) measurement and up projection to recover the model dimension, and iv) optimization through the parameter-shift rule. The following subsections provide details of each stage and outline its theoretical advantages.

\begin{figure}[t]
    \centering
    \includegraphics[width=0.75\columnwidth]{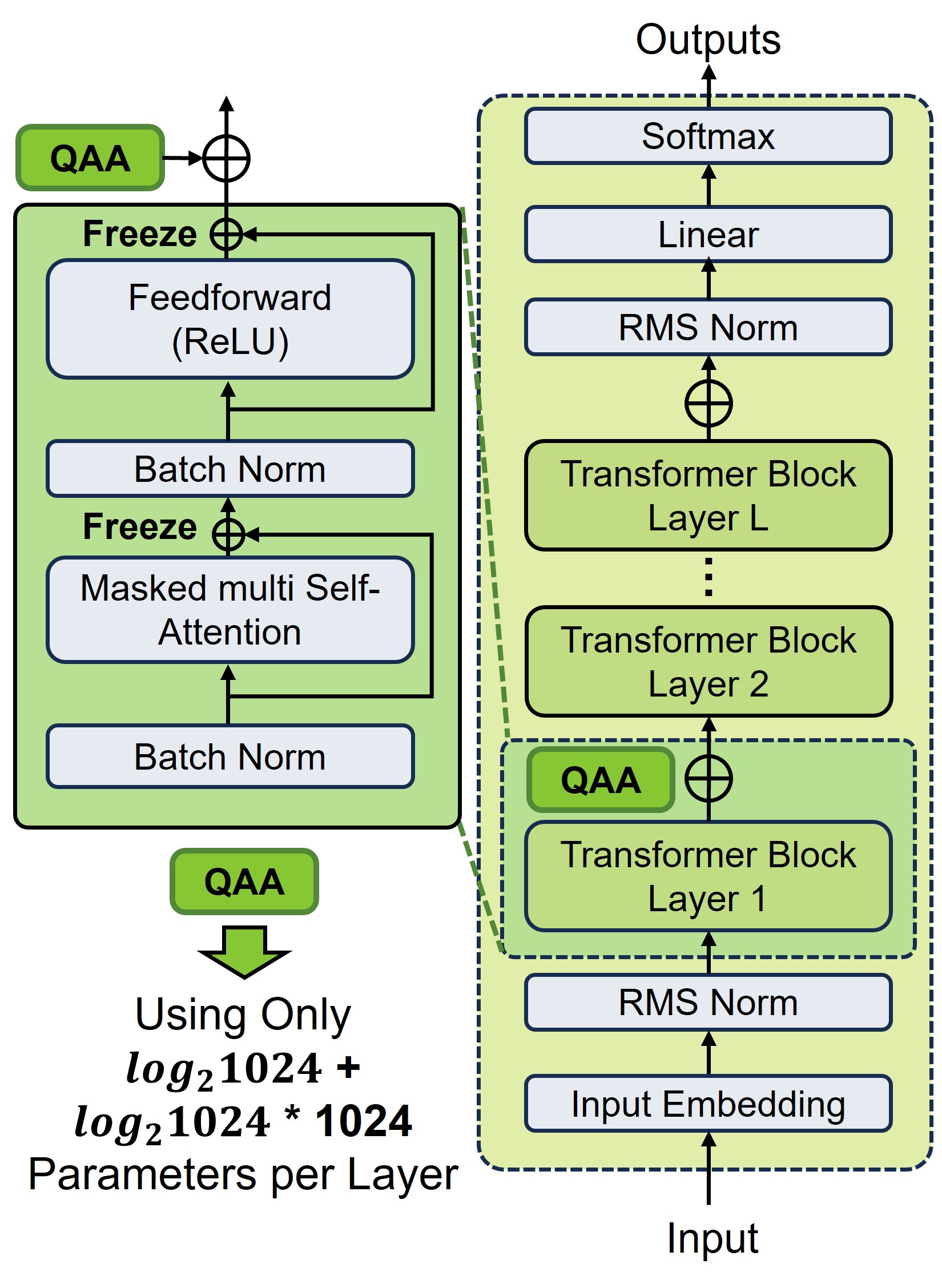}
    \caption{Illustration of how QAA operates within the GPT architecture.}
    \label{fig:2}
\end{figure}

\subsection{Quantum Amplitude Embedding}
A quantum state defines the configuration of a quantum system and is mathematically represented as a unit vector in a complex Hilbert space $\mathbb{C}^d$. Let $\{|i\rangle\}_{i=1}^{d}$ denote an orthonormal basis of $\mathbb{C}^d$, where each $|i\rangle$ corresponds to a distinct classical state. A general state is expressed as,
\begin{equation}
    |\psi\rangle = \sum_{i=1}^{d}\nolimits \alpha_i |i\rangle, \quad \text{with} \quad \sum_{i=1}^{d}\nolimits |\alpha_i|^2 = 1,
\end{equation}
where $\alpha_i \in \mathbb{C}$ are amplitudes. A measurement collapses $|\psi\rangle$ into basis state $|i\rangle$ with probability $|\alpha_i|^2$, thereby providing a probabilistic encoding of all classical indices in superposition. This property enables quantum systems to represent exponentially many configurations simultaneously~\cite{TMC_GSKIM}.

In QAA, hidden vectors from transformer layers are encoded into quantum states using amplitude embedding. Let $x \in \mathbb{R}^d$ denote a hidden activation vector. The smallest number of qubits $n$ is chosen such that $2^n \geq d$, embedding $x$ into an $n$-qubit Hilbert space $\mathbb{C}^{2^n}$. The vector is normalized as,
\begin{equation}
    \tilde{x} = \frac{x}{\|x\|_2},  
\end{equation}
where $\|x\|_2 = \sqrt{\sum_{k=0}^{d-1} x_k^2}$, and $x_k$ is the $k$-th entry of the vector $x$ and $\|x\|_2$ denotes the $\ell_2$ norm. This guarantees that $\tilde{x}$ has unit norm, ensuring physical validity as a quantum state. The normalized vector is mapped to,
\begin{equation}
    \ket{x} = \sum_{k=0}^{2^n - 1}\nolimits \tilde{x}_k \ket{k},
\end{equation}
where $\ket{k}$ denotes the computational basis state corresponding to the binary encoding of index $k$. This process compresses the $d$-dimensional vector into $\log_2 d$ qubits while preserving the structure of the original activations~\cite{QAA}.

\subsection{Parameterized Quantum Circuit}
After embedding, the quantum state transforms a PQC $U(\boldsymbol{\theta})$. A single-qubit gate rotates each qubit $j$,
\begin{equation}
    R_X(\theta_j) = \exp\left(-i \frac{\theta_j}{2} X\right),
\end{equation}
where $\theta_j \in \mathbb{R}$ is a trainable parameter and $X$ is the Pauli-X matrix $ \begin{bmatrix}0 & 1 \\ 1 & 0\end{bmatrix} $. These rotations introduce non-linear degrees of freedom. To capture correlations between qubits, CNOT gates are applied,
\begin{equation}
    \text{CNOT}_{j,j+1} \ket{a}_j \ket{b}_{j+1} = \ket{a}_j \ket{a \oplus b}_{j+1},
\end{equation}
where $a,b \in \{0,1\}$ and $\oplus$ denotes the XOR operation. This introduces quantum entanglement, which allows the PQC to model joint dependencies beyond local linear effects~\cite{ICASSP_FL_QNN}.

\subsection{Measurement and Up Projection}
The evolved quantum state is represented as $\ket{\psi(\boldsymbol{\theta})} = U(\boldsymbol{\theta}) \ket{x}$. To extract classical information, each qubit $j$ is measured in the Pauli-Z basis as,
\begin{equation}
    z_j = \bra{\psi(\boldsymbol{\theta})} Z_j \ket{\psi(\boldsymbol{\theta})}, \quad j=1,\dots,n,
\end{equation}
where $Z_j$ is the Pauli-Z observable acting on qubit $j$. This produces a vector $z \in \mathbb{R}^n$ that summarizes the circuit output. Since $n \ll d$, a linear up projection is applied as,
\begin{equation}
    \hat{x} = W^\top z, \quad W \in \mathbb{R}^{n \times d},
\end{equation}
where $W$ is a trainable projection matrix. The result $\hat{x}$ is interpreted as the residual update $\Delta h$, which is added to the frozen hidden state $h_{\text{base}}$, which forms the adapted representation $h_{\text{adapted}} = h_{\text{base}} + \Delta h$.

\subsection{Optimization with Parameter-Shift Rule}
To train the trainable parameters of PQC $\boldsymbol{\theta}$, QAA employs the parameter-shift rule. For an observable $\mathcal{O}$, the expectation value is defined as,
\begin{equation}
    f(\theta_j) = \bra{\psi(\boldsymbol{\theta})} \mathcal{O} \ket{\psi(\boldsymbol{\theta})}.
\end{equation}
Its gradient with respect to $\theta_j$ is computed as follows,
\begin{equation}
    \frac{\partial f}{\partial \theta_j} = \tfrac{1}{2} \Big[f(\theta_j + \tfrac{\pi}{2}) - f(\theta_j - \tfrac{\pi}{2})\Big].
\end{equation}
This avoids direct differentiation through non-analytic quantum operations. The gradients are combined with a classical loss $\mathcal{L}$, and each parameter is updated as,
\begin{equation}
    \theta_j \leftarrow \theta_j - \eta \cdot \frac{\partial \mathcal{L}}{\partial \theta_j},
\end{equation}
where $\eta$ is the learning rate. This hybrid procedure integrates quantum parameter updates into classical backpropagation.

\subsection{Implementation QAA on LLMs}
The integration of QAA into LLMs is designed to replace conventional adapter modules with quantum-enhanced components while keeping the majority of the backbone frozen~\cite{liu2022fewshot}. As illustrated in Fig.~\ref{fig:2}, QAA modules are inserted at multiple transformer layers, specifically after the self-attention and feedforward blocks. The base transformer weights remain fixed, and QAA generates task-specific residuals that are added to the hidden representations. This design enables efficient adaptation without modifying the full parameter set of the pre-trained model.
This implementation strategy highlights two key advantages. First, QAA enables scalable integration within LLMs by operating as a plug-in module, which ensures compatibility with transformer-based architectures. Second, it preserves the representational richness of hidden states through quantum-inspired transformations, which achieves expressive and efficient fine-tuning with logarithmic qubit complexity and linear projection overhead.

\begin{table}[t!]
\centering
\footnotesize
\caption{Specifications of hardware platforms, and software environments for Evaluation.}
\renewcommand{\arraystretch}{1.1}
\begin{tabular}{l||l} 
\Xhline{2\arrayrulewidth}
\textsf{\textbf{System}} & \textsf{\textbf{Specification (Value)}} \\ \midrule
Platform (PC) & OS: Ubuntu 20.04 \\
              & CPU: Intel(R) Xeon(R) CPU E5-2698 v4 \\
              & GPU: NVIDIA RTX-4090 (24 GB VRAM) \\ 
              & Memory: 256 GB DDR5 \\
\midrule
Software version & Python: v3.10 \\
                & CUDA: v11.8 \\
                & PyTorch: v2.1.2 \\
                & Transformers (HF): v4.44.2 \\
                & PEFT: v0.11.1 \\
                & Datasets: v2.14.5 \\
                & Pennylane: v0.36.0 \\ 
\bottomrule
\end{tabular}
\label{tab:2}
\end{table}

\section{Performance Evaluation}
To compare the representative PEFT methods, including full tuning, LoRA, SoRA, Prefix tuning, and the proposed QAA, experiments are conducted under the simulation environment summarized in Table~\ref{tab:2}.

\subsection{Quantitative Results}
Table~\ref{tab:3} reports the performance of various PEFT strategies in terms of BLEU, BERTScore (F1), and ROUGE metrics, where each value represents the average score computed over 100 generation sentences based on the Alpaca dataset. Full fine-tuning achieves the highest overall accuracy with BLEU of 12.19, BERTScore of 84.69, and ROUGE of 20.39/12.64/20.25, but at the cost of training all parameters. LoRA achieves competitive performance, with BLEU of 3.45 and BERTScore of 78.33, while requiring only 0.12\% of the parameters. SoRA further improves efficiency by adaptively reducing redundant ranks, which yields BLEU of 2.67 and BERTScore of 77.67 with 0.09\% parameters. Prefix tuning, despite using 0.44\% parameters, shows lower effectiveness with BLEU of 0.38 and BERTScore of 58.29, indicating difficulty in stable convergence for generative tasks. 
QAA demonstrates a strong balance between efficiency and performance. With only 0.09\% trainable parameters, it achieves BLEU of 2.96, BERTScore of 78.74, and ROUGE of 15.01/3.89/13.55. Although full fine-tuning remains the upper bound, QAA consistently outperforms Prefix tuning and shows comparable performance to LoRA and SoRA while maintaining a significantly smaller parameter budget. These results validate that QAA provides a promising path for efficient yet expressive LLM adaptation.

\begin{table}[t]
\centering
\caption{Comparison of the NLG evaluation metrics using different PEFT methods.}
\footnotesize
\begin{tabular}{c|c|c c c}
\Xhline{2\arrayrulewidth}
Method & \#TP Ratio & BLEU & BERTF1 & ROUGE \\
\hline
Full   & 100\%  & \textbf{12.19} & \textbf{84.69} & \textbf{20.39} / \textbf{12.64} / \textbf{20.25} \\
LoRA   & 0.12\% &  3.45 & 78.33 &  13.60 / 6.66 / 10.57 \\
SoRA   & 0.10\% &  0.67 & 77.67 &  7.43 / 1.43 / 5.41 \\
Prefix & 0.44\% &  0.38 & 58.29 &  7.18 / 1.82 / 6.77 \\
\textbf{QAA} & \textbf{0.09\%} & 2.96 & 78.74 & 15.01 / 3.89 / 13.55 \\
\Xhline{2\arrayrulewidth}
\end{tabular}
\label{tab:3}
\vspace{-3mm}
\end{table}

\begin{figure}[t]
    \centering
    \includegraphics[width=0.99\columnwidth]{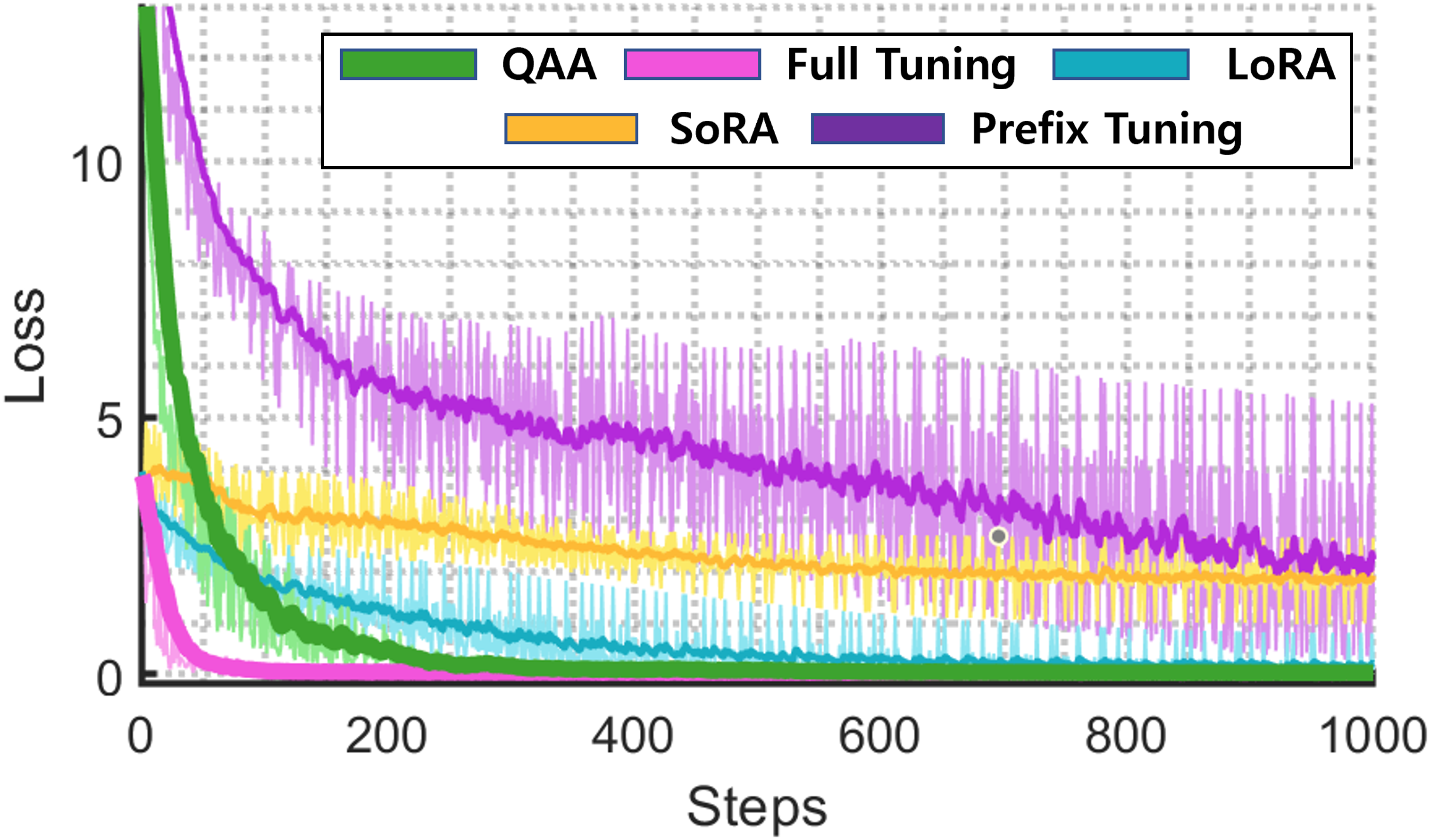}
    \caption{Training loss comparison across 1,000 steps.}
    \label{fig:3}
    \vspace{-3mm}
\end{figure}

\subsection{Training Loss Convergence Analysis}
The training loss curves across 1,000 steps are illustrated in Fig.~\ref{fig:3}. Full fine-tuning converges fastest due to the complete parameter space. Among PEFT methods, QAA exhibits a notably smooth and rapid convergence trajectory, outperforming Prefix and SoRA tuning and closely following LoRA. The variance in loss reduction for QAA remains lower than that of LoRA, SoRA, and Prefix tuning, which highlights the stabilizing effect of amplitude embedding and quantum circuit expressivity. These observations confirm that QAA provides stable gradient flow with reduced parameter complexity, enabling efficient training without sacrificing convergence speed.

\section{Conclusion}
This work provided a comprehensive survey and analysis of PEFT strategies for LLMs, including full tuning, LoRA, SoRA, Prefix tuning, and QAA. Through systematic evaluation, QAA is shown to deliver a favorable balance between efficiency and performance, which offers competitive performance with significantly fewer trainable parameters. The overall analysis highlights QAA as a promising direction that complements classical PEFT methods while demonstrating the potential of quantum deep learning in future LLM adaptation.

\begin{spacing}{0.6}
\footnotesize
\bibliographystyle{IEEEbib}
\bibliography{refs}
\end{spacing}

\end{document}